\documentclass[11pt,a4paper]{article}
\usepackage{jcappub}
\usepackage{mathrsfs}
\title{Implications of fast radio bursts for superconducting cosmic strings}

\author[a,b]{Yun-Wei Yu,}\emailAdd{yuyw@mail.ccnu.edu.cn}

\author[c]{Kwong-Sang Cheng,}%

\author[b,d]{Gary Shiu,}

\author[b]{Henry Tye}

\affiliation[a]{Institute of Astrophysics, Central China Normal
University, Wuhan, China}

\affiliation[b]{Institute for Advance Study, The Hong Kong
University of Science and Technology, Hong Kong, China}

\affiliation[c]{Department of Physics, The University of Hong Kong,
Hong Kong, China}

\affiliation[d]{Department of Physics, University of Wisconsin,
Madison, WI 53706, USA}

\abstract{Highly beamed, short-duration electromagnetic bursts could
be produced by superconducting cosmic string (SCS) loops oscillating
in cosmic magnetic fields. We demonstrated that the basic
characteristics of SCS bursts such as the electromagnetic frequency
and the energy release could be consistently exhibited in the
recently discovered fast radio bursts (FRBs). Moreover, it is first
showed that the redshift distribution of the FRBs can also be well
accounted for by the SCS burst model. Such agreements between the
FRBs and SCS bursts suggest that the FRBs could originate from SCS
bursts and thus they could provide an effective probe to study SCSs.
The obtained values of model parameters indicate that the loops
generating the FRBs have a small length scale and they are mostly
formed in the radiation-dominated cosmological epoch.}


\begin{document}

\maketitle

\flushbottom

\section{Introduction}
Cosmic strings, which have many small-scale wiggles, are formed as
linear topological defects during symmetry breaking phase transition
in the very early universe \cite{vil94}. As a result of string
interactions, a large number of closed loops could detach from the
string network and the length scale of the loops is comparable to
that of the string wiggles. In a wide class of grand unified models,
cosmic strings are predicted to behave as superconducting wires
\cite{Wi85}. Therefore, as a superconducting cosmic string (SCS)
moves through the cosmic magnetic fields of strength of $B$, it is
able to develop an electric current at a rate of $dI/dt\sim
(ce^2/\hbar)B$ \cite{chu86}, where $c$ is the speed of light, $e$
the electron charge, and $\hbar$ the Planck constant. Furthermore,
the superconducting loops oscillating in the magnetic fields can act
as an alternating current generator. Consequently, some highly
beamed, short electromagnetic (EM) bursts could be produced by the
loops at some special points (i.e., cusps where the speed of the
string segment is very close to $c$) \cite{vil87,bla01,Os86,spe87}.
Due to the EM and probably much stronger gravitational wave (GW)
radiations, the loops would shrink with time.

The EM bursts of SCS loops could provide a valuable probe to
discover SCSs. As relics from the early universe, the discovery of
SCSs would give insight into the physics of fundamental interactions
that governed cosmic evolution. Specifically, at very high
redshifts, the EM energy released from SCS bursts could be absorbed
by the dense surrounding medium to form a fireball, which could
subsequently generate a gamma-ray burst (GRB) through internal
dissipations \cite{Ba87,bre01,Pac88}. It is undoubtedly encouraged
to try finding signatures of SCSs from GRB observations
\cite{cheng10}. Unfortunately, the present sample of high-redshift
GRBs is very small \cite{Kis09} and, more seriously, it is not easy
to identify SCS-produced GRBs from typical GRBs originating from
collapsars and compact binary mergers \cite{cheng11}. In contrast,
at relatively low redshifts, the SCS burst emission could
successfully penetrate through the intergalactic medium (IGM) and be
detected by radio telescopes \cite{vac08, cai12}. Particularly, as
suggested by Vachaspati \cite{vac08}, such a radio transient signal
is very likely to have been reported by Lorimer et al. \cite{lor07}
in a survey with the 64-m Parkes radio telescope, because the basic
features of the Lorimer burst, if it has a cosmological distance,
can be reasonably explained by the SCS burst model \cite{vac08}.

Very recently, after the High Time Resolution Universe (HTRU) survey
with the Parkes telescope, Thornton et al. \cite{tho13} reported
four new Lorimer burst-like radio transients (presently called fast
radio bursts; FRBs), the parameters of which are listed in Table I.
First of all, as claimed by Thornton et al., the anomalously high
dispersion measures (DMs) of all four FRBs coupled with their high
Galactic latitudes confirm the cosmological origin with a redshift
$z\sim0.5-1$. Secondly, their basic properties are identical to the
Lorimer burst and thus these new Lorimer burst-like events provide
further support of the consistency between the FRBs and SCS bursts.
Finally and most importantly, the accumulated number of FRBs could
make it more stringent to constrain the event rate of SCS bursts and
even its redshift evolution. Although the present sample is not
large, such an attempt may still effectively substantiate the role
of the FRBs as observational signatures of SCSs.

\section{SCS burst model}
\subsection{Basic characteristics of the bursts}
For a length of a SCS loop of $l$ at redshift $z$, the duration of
the loop transient radiation could in principle be determined by the
period of the loop oscillation as $T_{l}\sim l/c$, if the moving
velocity is subrelativistic. However, the closer to the cusp, the
higher the speed of the string segments. Therefore, as analyzed by
Babul \& Paczy{\'n}ski \cite{Ba87}, the duration of the EM burst for
an observer should be corrected to $\Delta t_{\rm burst}\sim
f_zT_l/\gamma^3$, where $f_z\equiv (1+z)$ is introduced due to the
cosmological time dilation and $\gamma$ is the Lorentz factor of the
string segments near the cusp.

Following Reference \cite{vil87}, the power of the EM radiation of a
subrelativistically oscillating loop can be calculated with the use
of the magnetic dipole radiation formula as $P_0\sim {m^2\omega^4/
c^3}\sim {I^2/ c}$, where $m\sim Il^2/c$ is the magnetic moment of
the loop and $\omega\sim 1/T_l$ is the oscillation frequency. In
terms of the string tension $\mu$ (i.e., mass per unit length of the
string),  the maximum value of the current is found to be
$I_{\max}\sim\mu^{1/2} c^2$ which follows from the equation
$I_{\max}^2/c\sim\mu l c^2/T_{l}$. Then, the highest possible
Lorentz factor of the cusp can be determined by
$\gamma_{\max}=I_{\max}/I$ \cite{bla01b}. For a string segment
moving at a Lorentz factor of $\gamma$, its energy release will be
boosted by the Lorentz factor and be beamed within an angle of
$\theta\sim\gamma^{-1}$. Therefore, the angular distribution of the
energy release of a SCS burst can be written as \cite{vil87,bla01}
\begin{eqnarray}
{dE\over d\Omega}&\sim& {k_{\rm em}\gamma (P_0T_l)\over
\theta^2}\sim {k_{\rm em}I^2l\over c^2\theta^{3}} ,\rm ~for
~\theta>\theta_{\rm c}, \label{energy}
\end{eqnarray}
where the numerical coefficient $k_{\rm em}\sim 10$
\cite{vil87,bre01} and $\theta_{\rm c}\sim \gamma_{\max}^{-1}$. Here
$\theta=0$ is defined at the direction of the string motion. For an
observer at the light of sight of $\theta$, who cannot see the
radiation from the segments with $\gamma>\theta^{-1}$, the
isotropically-equivalent energy release can be written as
\begin{eqnarray}
E_{\rm iso}=4\pi {dE\over d\Omega}\sim 4\pi k_{\rm em}{I^2l\over
c^2\theta^{3}}.\label{Eiso}
\end{eqnarray}
This result could usually be much higher than the real total energy
release of the SCS burst as $E_{\rm tot}\sim k_{\rm
em}I^2l/(c^2\theta_{\rm c})\sim k_{\rm em}I\mu^{1/2}l$
\cite{bre01,vil87,spe87}, which is obtained by integrating Eq.
(\ref{energy}) over the whole solid angle.

Finally, the observational frequency of the EM burst can be
estimated by \cite{bla01,cai12}
\begin{eqnarray}
\nu\sim{1\over  f_z}{c\over\theta^{3}l}\sim {1\over \Delta t_{\rm
burst}},\label{nu}
\end{eqnarray}
which indicates that the intrinsic duration of SCS bursts can be
simply inferred from the observational frequency.

\subsection{Loop density}
The timescale of the loop shrinkage due to the EM radiation can be
estimated by $\tau_{\rm em}\sim({\mu l c^2/ E_{\rm
tot}})T_l=\mu^{1/2}lc/(k_{\rm em}I)$. In contrast, the shrinking
timescale due to GW radiation can be written as $\tau_{\rm gw}\sim{l
c/ (k_{\rm gw}G\mu)}$ with a numerical coefficient $k_{\rm gw}\sim
50$ \cite{bre01}, where $G$ is the Newton's gravitational constant.
Comparing $\tau_{\rm em}$ with $\tau_{\rm gw}$, we can get a
critical current of $I_{*}=(k_{\rm gw}/k_{\rm
em})G\mu^{3/2}=1\times10^{19}\mu_{17}^{3/2}\rm~ esu~s^{-1}$, below
which the loop shrinkage is dominated by the GW radiation. Hereafter
the conventional notation $Q_x=Q/10^{x}$ is adopted in the cgs
units. Denoting the shrinking rate of the loop by $\Gamma\equiv
-dl/dt$, we have $\Gamma=\Gamma_{\rm em}+\Gamma_{\rm
gw}=l({\tau_{\rm em}}^{-1}+{ \tau_{\rm gw}}^{-1})$. Then for a SCS
loop having a length of $l$ at redshift $z$, its initial length
before the radiation shrinkage is given by \cite{vac08}
\begin{eqnarray}
l_{\rm i}=l(z)+\Gamma\left[t(z)-t_{\rm i}\right],\label{lini}
\end{eqnarray}
where the birth-time of the loop, $t_{\rm i}$, is usually much
smaller than $t(z)$ of interest here.

Numerical simulations showed that the string network scales with the
horizon, i.e., the typical curvature radius of long SCSs and the
distance between them are both on the order of the horizon size
\cite{ben90}. Therefore, the differential density of the SCS loops
as a function of their initial length can be written as ${dn/
dl_{\rm i}} \sim {\left[l_{\rm i}^{5/2} (ct)^{3/2}\right]^{-1}}$ and
${dn/ dl_{\rm i}} \sim {(l_{\rm i}ct)^{-2}}$, for the loops formed
in the radiation- and matter-dominated cosmological epochs,
respectively \cite{cai12,bra86}. More specifically, by considering
of the contribution of the loops survived from the
radiation-dominated era, the distribution function in the
matter-dominated era of interest here should be taken as follows
\cite{cai12}:
\begin{equation}
{dn\over dl_{\rm i}} \sim  \left(1+\sqrt{ct_{\rm eq}\over l_{\rm
i}}\right)\frac{1}{l_{\rm i}^{2} (ct)^{2}},\label{density}
\end{equation}
where $t_{\rm eq}\sim 2\times10^{12}$ s is the time of
radiation-matter equality.

\begin{table*}
\centering \caption{Observational parameters for the FRBs$^\dag$
\cite{tho13}}
\begin{tabular}{ c  c c c c c}
\hline \hline
   FRBs  & $\Delta t_{\rm obs}$ (ms)  & $S_{\nu}$ (Jy) &  DM ($\rm cm^{-3}pc$) & $z$ &  $E_{\rm iso}$ ($10^{40}$erg) \\
 \hline
  110220 & 5.6$\pm$0.1&  1.3      & 944.38$\pm$0.05  & 0.81   & 5.0  \\
  110627 & $<$1.4     &  0.4      & 723.0$\pm$0.3    & 0.61   & $<$0.2  \\
  110703 & $<$4.3     &  0.5      & 1103.6$\pm$0.7   & 0.96   & $<$2.1  \\
  120127 & $<$1.1     &  0.5      & 553.3$\pm$0.3    & 0.45   & $<$0.1  \\
\hline
\end{tabular}\\
$^\dag$The widths and fluxes are all measured at the frequency of
1.3 GHz.
\end{table*}

\section{Implications from FRBs}
\subsection{Constraining the length of loops}
Eq. (\ref{nu}) indicates that the intrinsic duration of SCS bursts
at the frequency $\nu\sim1$ GHz is extremely short, i.e., $\Delta
t_{\rm burst}\sim\nu^{-1}\sim 10^{-9}$ s, which is dramatically
shorter than the observed widths of FRBs. This is because the
observed duration of radio transient emission can be significantly
influenced by the scattering by the turbulent IGM and the time
resolution of telescopes. Therefore, we have \cite{miy13}
\begin{eqnarray}
\Delta t_{\rm obs}=\max\left\{\left(\Delta t_{\rm burst}^2+\Delta
t_{\rm scat}^2\right)^{1/2},\Delta t_{\rm res}\right\}.
\end{eqnarray}
In view of the time resolution of $\Delta t_{\rm res}\sim 64\rm\mu
s$ of the HTRU survey \cite{kei10}, the observed duration of FRB
110220 as $\Delta t_{\rm obs}=5.6$ ms indicates that the value of
$\Delta t_{\rm scat}$ is probably on the order of milliseconds and
$\Delta t_{\rm obs}\approx \Delta t_{\rm scat}$. In addition,
$\Delta t_{\rm scat}$ is theoretically considered to evolve with
redshift (e.g., in References \cite{vac08,cai12}), which however has
not been exhibited in the observational sample due to its small
size.

In any case, the isotropically-equivalent energy release of the
observed FRBs can be calculated by
\begin{eqnarray}
E_{\rm iso}\approx 4\pi d_{c}^2 \Delta t_{\rm obs}\Delta\nu
S_{\nu}f_z,
\end{eqnarray}
where the frequency bandwidth is taken to $\Delta\nu=0.4$ GHz
\cite{tho13} and the values of the duration $\Delta t_{\rm obs}$ and
the flux density $S_{\nu}$ are listed in Table I. As shown in the
last column of Table I, the calculated energies are on the order of
magnitude of $\sim10^{39}-10^{41}$ erg. Here the comoving distances
of the sources
$d_c=(c/H_0)\int_0^{z}(f_{z'}^3\Omega_m+\Omega_\Lambda)^{-1/2}dz'$
can be calculated with the redshifts derived from the measured DMs.
The cosmological parameters read $H_0=71\rm km~s^{-1}Mpc^{-1}$,
$\Omega_m=0.27$, and $\Omega_{\Lambda}=0.73$.

By attributing the observed FRBs to the EM bursts of SCS loops, the
parameters of the SCS burst model can be constrained. To be
specific, by taking $\nu\sim 1$ GHz and $E_{\rm iso}\sim 10^{40}$
erg as reference values and adopting the following relationship
\cite{chu86}
\begin{eqnarray}
I\sim (e^2/\hbar)Bl,
\end{eqnarray}
we can solve for the length of the SCS loops from Eqs. (\ref{Eiso})
 and (\ref{nu}) as
\begin{eqnarray}
l\sim 8\times10^{13}f_z^{-5/4} B_{0,-6}^{-1/2}{E_{\rm
iso,40}^{1/4}\nu_9^{-1/4}}\rm~ cm.\label{length}
\end{eqnarray}
The above value is about $10^3$ times longer than that found by
Vachaspati \cite{vac08}\footnote{In Reference \cite{vac08}, the
intrinsic duration of SCS bursts is incorrectly overestimated by a
factor of $\gamma^2$ (see \cite{cai12}), which leads to the
underestimation of $l$.}, but it is still very small on a
cosmological scale. Here the cosmic magnetic fields are assumed to
be frozen in the cosmic plasma, at least, for relatively low
redshfits. This yields $B(z)=B_0f_z^2$. However, more complicatedly,
the fields probably distribute inhomogeneously and the field
strength could vary on different field coherent lengths
\cite{blasi99}. Here an upper limit value of the present strength
$B_0\sim1\mu$G is adopted self-consistently corresponding to the
short length of the loops .

The current on the loops can be derived from Eq. (\ref{length}),
\begin{eqnarray}
I\sim 2\times10^{16}f_z^{3/4} B_{0,-6}^{1/2}{E_{\rm
iso,40}^{1/4}\nu_9^{-1/4}}\rm esu~s^{-1},
\end{eqnarray}
which is much lower than the critical current $I_*$ except for a
very small $\mu$. Strictly speaking, on one hand, the above current
could be increased by an increasing loop length. On the other hand,
however, the magnetic fields on longer length scales and farther
away from galaxy clusters could become much lower, e.g.
$10^{-9}-10^{-8}$ G \cite{blasi99}. Hence, the increased current
could still not exceed $I_*$. In the following calculations, the
shrinkage of the loops is considered to be dominated by the GW
radiation. This yields
\begin{eqnarray}
l_{\rm i}\sim\Gamma_{\rm gw}t\sim{k_{\rm gw}G\mu\over c}t\sim
5\times10^{18}f_{z}^{-3/2}\mu_{17}\rm ~cm,\label{dlgw}
\end{eqnarray}
where, for an analytical expression, the time is approximated by
$t(z)\approx(1/H_0)f_z^{-3/2}$. It should be noted that the redshift
here corresponds to the FRB generation but not to the loop
formation.

\subsection{Fitting to the accumulated numbers of FRBs}
Following Eqs. (\ref{lini}) and (\ref{density}), the observational
burst rate of the SCS loops of a length $l$ at redshift $z$ can be
written as
\begin{eqnarray}
\dot{R}(z)\sim{\theta^2\over  4 T_{l}}\int_{l+\Gamma t}{dn\over
dl'_{\rm i}} dl'_{\rm i}
\sim{\theta^2c\over  4l}\left(1+{2\over3}\sqrt{ct_{\rm eq}\over
l+\Gamma t}\right)\frac{1}{\left(l+\Gamma
t\right)\left(ct\right)^{2}}.
\end{eqnarray}
By substituting Eqs. (\ref{length}) and (\ref{dlgw}) into the above
equation and considering  $l\ll l_{\rm i}\sim\Gamma_{\rm gw}t\ll
ct_{\rm eq}$, we can approximate the observational burst rate by
\begin{eqnarray}
\dot{R}(z)\sim{\theta^2t_{\rm eq}^{1/2}\over
6 c^{1/2}l\Gamma_{\rm gw}^{3/2}t^{7/2}}
\sim4\times10^4{ f_z^{20/3}B_{0,-6}^{5/6}\mu_{17}^{-3/2}E_{\rm
iso,40}^{-5/12}\nu_9^{-1/4}}~\rm Gpc^{-3}yr^{-1}, \label{rate1}
\end{eqnarray}
where the viewing angle reads $\theta\sim(\nu f_zl/c)^{-1/3}\sim
7\times10^{-5}f_z^{1/12} B_{0,-6}^{1/6}E_{\rm
iso,40}^{-1/12}\nu_9^{-1/4}$.

Specifically, for the HTRU survey with the Parkes telescope, the
observational threshold at $\nu=1.3$ GHz with a bandwidth $\Delta
\nu=0.4$ GHz can be estimated by
\begin{eqnarray}
E_{\rm iso,th}=4\pi d_{c}^2 \Delta t_{\rm
obs}\Delta\nu S_{\nu,\rm th}f_z
\approx2\times10^{40}(f_z^{1/2}-1)^2\rm erg,\label{Eth}
\end{eqnarray}
where the comoving distance is approximated analytically by
$d_c\approx(3c/H_0)f_z^{-1/2}(f_z^{1/2}-1)$ \cite{bre01}, $\Delta
t_{\rm obs}\approx\Delta t_{\rm scat}\sim 1$ ms, and the flux
sensitivity is taken to be $S_{\nu,\rm th}=0.3$ Jy as a reference
value\footnote{The determination of $S_{\nu,\rm th}$ actually is not
trivial, which depends on the sky region, the time resolution, and
the DM etc. Miyamoto et al. \cite{miy13} suggested a value of
$S_{\nu,\rm th}=0.61$ mJy for the HTRU survey by following Reference
\cite{kei10}, which, however, is appropriate for a pulsar survey but
may not be for a single-pulse search. Here we take the reference
value of $S_{\nu,\rm th}$ according to Fig. 3 in Reference
\cite{bur11} with a rebinned time resolution of $0.512$ ms
\cite{tho13}.}. Substituting Eq. (\ref{Eth}) into (\ref{rate1}), we
can get the event rate of SCS bursts for the HTRU single-pulse
search as
\begin{eqnarray}
\dot{R}_{\rm HTRU}(z)\sim3\times10^4f_z^{20/3}(f_z^{1/2}-1)^{-5/6}
 B_{0,-6}^{5/6}\mu_{17}^{-3/2}~\rm Gpc^{-3}yr^{-1},
\end{eqnarray}
Then the observed accumulated number of SCS bursts can be calculated
by
\begin{eqnarray}
N(\leq z)={\mathcal T}{\mathcal A\over 4\pi} \int_0^z
{\dot{R}_{\rm HTRU}(z')\over f_{z'}}{dV_p(z')}
\sim311
B_{0,-6}^{5/6}\mu_{17}^{-3/2}\int_1^{f_z}{\left[x(x^{1/2}-1)^{7}\right]^{1/6}
}dx, \label{number}
\end{eqnarray}
where $\mathcal T=270$ s is the duration of each pointing
observation, $\mathcal A=4500\rm deg^2=1.4~sr$ is the area of the
survey \cite{tho13}, the factor $f_{z'}$ is due to the cosmological
time dilation of the observed rate, and the proper volume element is
given by $dV_p\approx 54\pi (c/H_0)^3f_z^{-11/2}(f_z^{1/2}-1)^2dz$
\cite{bre01}.

\begin{figure}
\centering\resizebox{0.6\hsize}{!}{\includegraphics{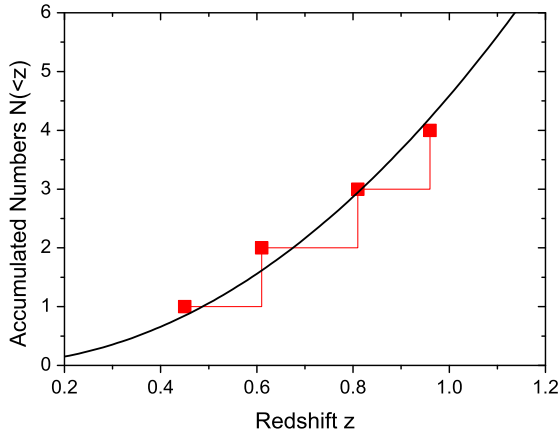}}
\caption{Fitting to the redshift distribution of the observed FRBs
(solid squares) by the SCS burst model (Eq. \ref{number}; solid
line).}
\end{figure}

Finally, in Fig. 1 we present the accumulated numbers of FRBs by the
solid squares. In order to avoid the complicacy due to the different
telescope parameters in different surveys, here we only invoke the
Thornton et al.'s data \cite{tho13}, but exclude FRB 010824
\cite{lor07}, FRB 010621 \cite{kea12}, and the first non-Parkes FRB
121102 discovered in the 1.4-GHz Pulsar ALFA survey with the Arecibo
Observatory \cite{spi14}. The best fitting to the data by Eq.
(\ref{number}) is shown by the solid line, which
correspomds to
\begin{eqnarray}
\mu\sim 5.5\times10^{17}B_{0,-6}^{5/9}{\rm ~g~cm^{-1}}
  \sim 6.1 \times 10^{27}B_{0,-6}^{5/9} \rm ~GeV^{-2}. \nonumber \\
  \label{tension}
\end{eqnarray}
Such a result is typical and well consistent with some previous
cosmological and astrophysical constraints on SCSs \cite{miy13}.
With the above result, the previously used condition of $l\ll l_{\rm
i}\sim\Gamma_{\rm gw}t\ll ct_{\rm eq}$ can be confirmed. In such a
case, it can be known that the loops responsible for the FRBs are
mostly formed in the radiation-dominated era. Moreover, the
comparison between $l_{\rm i}$ and $ct_{\rm eq}$ indicates that the
ratio of $\alpha=l_{\rm i}/(ct_i)$ is higher than $\sim3\times
10^{-4}$.

\section{Conclusion and discussions}
By ascribing the observed FRBs to SCS EM bursts and using the EM
frequency, duration, energy release, and number of the FRBs, we
constrain the most important parameters of the SCS burst model such
as $l(z)$, $l_{\rm i}$, and $\mu$. The obtained typical values of
the parameters reconfirm the possible connection between the FRBs
and SCS bursts. More importantly, we first investigate the redshift
distribution of the FRBs, both the normalization and the profile of
which are found to be well accounted for by the SCS burst model.
Such an excellent consistency provides a new and substantial
evidence for the role of FRBs as observational signatures of SCSs.
Furthermore, as implied by our results, the observed FRBs are
probably associated with the loops formed in the radiation-dominated
cosmological era.

An open issue remains as to why no high-redshift FRB has been
detected. On the contrary, the model prediction is that about 6 and
9 FRBs would appear within the redshift ranges of $1\leq z\leq1.5$
and $1.5\leq z\leq2$, respectively, if the event rate of SCS bursts
monotonously increase with an increasing redshift as shown. Here we
note that this contradiction could be resolved if there exist some
suppression effects at relatively high redshifts. Firstly, such a
suppression could arise from the decrease of the magnetic fields
surrounding SCS loops. On one hand, the cosmic fields are usually
considered to be amplified at a certain redshift, above which the
frozen field assumption could not be extended to. On the other hand,
SCS loops could be continuously captured and accreted by growing
matter perturbations. Therefore, at earlier times, the loops could
be much farther away from galaxy clusters, where the diffuse
magnetic fields are weaker. Secondly, there could be a cutoff on the
loop density at a certain short $l_{\rm i}$. In other words, above a
certain redshift, the shrinking timescale of all SCS loops could be
longer than the age of the universe at that time. Hence, above that
redshift, there is no SCS loop short enough to generate FRBs.
Thirdly, it could be easier to absorb FRBs by the relatively denser
IGM at higher redshifts. Finally, in any case, it also cannot be
ruled out that the absence of high-redshift FRBs could be partially
caused by some observational selection effects in the relevant range
of flux and duration or by the statistical bias due to the extremely
small size of the present data sample.

Besides the SCS burst explanation for the FRBs, many astrophysical
scenarios have also been proposed to account for the energy scale of
$\sim10^{40} \rm erg$ and the millisecond duration, e.g.,
hyperflares of soft gamma-ray repeaters \cite{pop07}, collapses of
super-massive neutron stars (NSs) to black holes at
several-thousands to million years old \cite{fal14} or at its baby
time \cite{zhang14}, mergers of double NSs \cite{tot13} or binary
white dwarfs \cite{kas13}. However, all of these astrophysical
models have not been confronted with the redshift distribution of
the FRBs. Additionally, some counterparts in other EM energy bands
could usually be predicted by most astrophysical models, which
however do not exist in the SCS burst model. At present, there is
indeed no counterpart reported to be associated with the FRBs,
although this could just be caused by the low angular resolutions of
the radio surveys. Another interesting point to note is that SCS
loops could cluster and form a halo about the galaxies
\cite{Chernoff:2009tp}. Thus if FRBs originate from the SCS loops,
we could expect to find some associated galaxies nearby the FRBs.
Moreover, a characteristic anisotropic distribution could appear,
when more FRB events are accumulated, e.g., through the surveys of
future radio telescopes such as SKA. This could provide some clues
to distinguish the SCS burst model from the other FRB models. Of
course, a larger sample of FRB events is also crucial for confirming
that their redshift distribution matches well with the SCS burst
model.

If the observed FRBs indeed originated from SCS bursts, one would
expect bursts of other light degrees of freedom such as GW bursts,
and possibly neutrino bursts to be radiated from cusps and kinks of
SCSs as well. Interestingly, with the cosmic string tension of the
order of $G \mu/c^2 \sim 4.1 \times 10^{-11} B_{0,-6}^{5/9}$
inferred from the FRBs, GW bursts radiated from SCS loops may be
detectable by the planned GW detectors such as LIGO/VIRGO and LISA
\cite{Da}. Combining data from the various channels could enable us
to distinguish between different types of cosmic strings. For
example, while SCSs arise commonly in particle physics models of the
early universe \cite{Wi85}, cosmic strings produced in string theory
models (see, e.g.,
\cite{Shiu:2002cb,Polchinski:2004ia,HenryTye:2006uv} for some recent
reviews) generically couple to the Standard Model degrees of freedom
with only gravitational strength \cite{Polchinski:2004ia}. This is
because gauge fields in string theory (typically) arise from
``branes" and stability requires the cosmic strings to be separated
from (most of) them. Models in which the cosmic superstrings couple
more strongly (with gauge interaction strength) to the EM waves can
be constructed but they constitute special cases. Turning this
around, these observational probes taken together can yield valuable
insights into the fundamental interactions of Nature. We hope to
return to the above issues in future work.

\section*{Acknowledgements}
This work is supported in part by the CRF Grants of the Government
of the Hong Kong SAR under HUKST4/CRF/13G. YWY is also supported by
the National Natural Science Foundation of China (grant No.
11473008) and the Program for New Century Excellent Talents in
University (grant No. NCET-13-0822). GS is additionally supported by
the DOE grant DE-FG-02-95ER40896.


\begin{thebibliography}{99}
\bibitem[\protect\citeauthoryear{}{}]{vil94}A. Vilenkin and E.P.S. Shellard, Cosmic Strings and Other
Topological Defects (Cambridge University Press, Cambridge, England,
1994).

\bibitem[\protect\citeauthoryear{}{}]{Wi85} E. Witten, Nucl. Phys. {\bf B249}, 557 (1985).

\bibitem[\protect\citeauthoryear{}{}]{chu86} E. M. Chudnovsky, G. B. Field, D. N. Spergel, and A. Vilenkin, Phys. Rev. {\bf D34}, 944
(1986).

\bibitem[\protect\citeauthoryear{}{}]{Os86} J. P. Ostriker, C. Thompson and E. Witten, Phys. Lett. {\bf
B180}, 231 (1986)

\bibitem[\protect\citeauthoryear{}{}]{vil87}A. Vilenkin and T. Vachaspati, Phys. Rev. Lett. {\bf 58}, 1041
(1987)

\bibitem[\protect\citeauthoryear{}{}]{spe87}D. N. Spergel, T. Piran, and J. Goodman, Nucl. Phys. {\bf B291}, 847
(1987).

\bibitem[\protect\citeauthoryear{}{}]{bla01}J. J. Blanco-Pillado, and K. D. Olum Nucl. Phys. {\bf B599}, 435 (2001).

\bibitem[\protect\citeauthoryear{}{}]{Ba87} A. Babul, B. Paczynski, and D. Spergel, Astrophys. J. {\bf 316},  L49
(1987); B. Paczynski, Astrophys. J. {\bf 335}, 525 (1988)

\bibitem[\protect\citeauthoryear{}{}]{Pac88}R. H.
Brandenberger, A. T. Sornborger, and M. Trodden, Phys. Rev. {\bf
D48},  940 (1993);  R. Plaga, Astrophys. J. {\bf 424}, L9 (1994);

\bibitem[\protect\citeauthoryear{}{}]{bre01}  V. Berezinsky,  B. Hnatyk, and A. Vilenkin, Phys. Rev. {\bf D64},
043004  (2001); V. Berezinsky,  B. Hnatyk, and A. Vilenkin, Baltic
Astronomy {\bf 13}, 289 (2004).

\bibitem[\protect\citeauthoryear{}{}]{cheng10}K. S. Cheng, Y. W. Yu, and T. Harko, Phys. Rev.
Lett. {\bf 104}, 231102 (2010)

\bibitem[\protect\citeauthoryear{}{}]{Kis09}  M. D. Kistler et al. Astrophys. J. {\bf 705},  L104
(2009); F. Y. Wang, Z. G. Dai, Astrophys. J., {\bf 727}, L34 (2011);
F. Y. Wang Astron. \& Astrophys. {\bf 556}, A90 (2013); F. Y. Wang,
Z. G. Dai, Astrophys. J. Supp., {\bf 213}, 15, (2014)

\bibitem[\protect\citeauthoryear{}{}]{cheng11}K. S. Cheng, Y. W. Yu, and T. Harko, Phys. Rev.
Lett. {\bf 106}, 259002 (2011)

\bibitem[\protect\citeauthoryear{}{}]{vac08} T. Vachaspati, Phys. Rev. Lett. {\bf 101}, 141301 (2008).

\bibitem[\protect\citeauthoryear{}{}]{cai12}Y. F. Cai, E. Sabancilar, and T.
Vachaspati, Phys. Rev. {\bf D85}, 023530 (2012); Y. F. Cai, E.
Sabancilar, D. A. Steer, and T. Vachaspati, Phys. Rev. {\bf D86},
043521 (2012).

\bibitem[\protect\citeauthoryear{}{}]{lor07}D. R. Lorimer, M. Bailes, M. A. McLaughlin, D. J.
Narkevic, and F. Crawford, Science {\bf 318}, 777 (2007).

\bibitem[\protect\citeauthoryear{}{}]{tho13}
{Thornton}, D., {Stappers}, B., {Bailes}, M., {et~al.}, Science,
{\bf 341}, 53 (2013)

\bibitem[\protect\citeauthoryear{}{}]{bla01b}J. J. Blanco-Pillado, K. D. Olum, A. Vilenkin, Phys.Rev. {\bf D63}, 103513 (2001)


\bibitem[\protect\citeauthoryear{}{}]{ben90}D. P. Bennett and F. R. Bouchet, Phys. Rev. {\bf D41}, 2408
(1990); B. Allen and E. P. S.Shellard, Phys. Rev. Lett. {\bf 64},
119 (1990); C. J. A. P. Martins and E. P. S. Shellard, Phys. Rev.
{\bf D73}, 043515 (2006); C. Ringeval, M. Sakellariadou, and F.
Bouchet, J. Cosmol. Astropart. Phys. {\bf 02}, 023 (2007); J. J.
Blanco-Pillado, K. D. Olum, and B. Shlaer, J. Comput. Phys. {\bf
231}, 98 (2012).

\bibitem[\protect\citeauthoryear{}{}]{bra86}R. H. Brandenberger and N. Turokf, Phys.Rev. {\bf D33}, 2182 (1986)

\bibitem[\protect\citeauthoryear{}{}]{blasi99}P. Blasi, S. Burles, and A. V. Olinto, Astrophys. J, {\bf 514}, L79,
(1999)

\bibitem[\protect\citeauthoryear{}{}]{miy13}K. Miyamoto, and K. Nakayama, J. of Cosmo. and
Astropart. Phys., {\bf 07}, 012 (2013)


\bibitem[\protect\citeauthoryear{}{}]{kei10}M. J. Keith, A. Jameson, W. van Straten, et al. MNRAS, {\bf 409},
619 (2010)

\bibitem[\protect\citeauthoryear{}{}]{bur11}S. Burke-Spolaor, M. Bailes, S. Johnston, et al., MNRAS, {\bf 416},
2465 (2011)

\bibitem[\protect\citeauthoryear{}{}]{kea12}Keane, E. F., Stappers, B. W., Kramer, M, Lyne,
A. G. MNRAS, {\bf 425}, L71 (2012)

\bibitem[\protect\citeauthoryear{}{}]{spi14}L. G. Spitler, J. M. Cordes, J. W. T. Hessels, et al.
Astrophys. J. {\bf 790}, 101 (2014)


\bibitem[\protect\citeauthoryear{}{}]{pop07}
S.~B. {Popov}, \& K.~A. {Postnov}, ArXiv e-prints: 1307.4924 (2013)

\bibitem[\protect\citeauthoryear{}{}]{fal14} H. Falcke, \& L. Rezzolla, A\&A, {\bf 562}, A137 (2014)

\bibitem[\protect\citeauthoryear{}{}]{zhang14}Zhang, B. 2014, ApJL, 780, L21

\bibitem[\protect\citeauthoryear{}{}]{tot13}
T. {Totani}, PASJ, {\bf 65}, L12 (2013)

\bibitem[\protect\citeauthoryear{}{}]{kas13}
 K. {Kashiyama}, K. {Ioka}, \& P. {M{\'e}sz{\'a}ros}, Astrophys.
J. Lett., {\bf 776}, L39 (2013)

\bibitem[\protect\citeauthoryear{}{}]{Da} T. Damour and A. Vilenkin, Phys. Rev. Lett. {\bf 85},
3761 (2000); T. Damour and A. Vilenkin, Phys. Rev. {\bf D 64},
064008 (2001); T. Damour and A. Vilenkin, Phys. Rev. {\bf D71},
063510 (2005).


\bibitem[\protect\citeauthoryear{}{}]{Shiu:2002cb}
  G.~Shiu,
  hep-th/0210313,
  prepared for the
{\it Proceedings of the First International
 Conference on String Phenomenology}, published by World Scientific (2003).

\bibitem[\protect\citeauthoryear{}{}]{Polchinski:2004ia}
J.~Polchinski,
  hep-th/0412244.

\bibitem[\protect\citeauthoryear{}{}]{HenryTye:2006uv}
  S.-H.~Henry Tye,
  Lect.\ Notes Phys.\  {\bf 737}, 949 (2008)
  [hep-th/0610221].


\bibitem[\protect\citeauthoryear{}{}]{Chernoff:2009tp}
D.~F.~Chernoff \& S.-H. Henry Tye, 2007, arXiv: 0709.1139;
D.~F.~Chernoff,
  arXiv: 0908.4077 [astro-ph.CO].











\end{thebibliography}
\end{document}